# Situation-based memory in spiking neuron-astrocyte network

Susanna Gordleeva[1,2,*], Yuliya A. Tsybina[1,3], Mikhail I. Krivonosov[1], Ivan Y. Tyukin[1,6], Victor B. Kazantsev[1,2,5], Alexey A. Zaikin[1,3,4] & Alexander N. Gorban[1,6]

[1] Lobachevsky State University of Nizhny Novgorod, 23 Gagarin Ave., 603022 Nizhny Novgorod, Russia

[2] Neuroscience and Cognitive Technology Laboratory, Innopolis University, 1 Universitetskaya Str., 420500 Innopolis, Russia

[3] Centre for Analysis of Complex Systems, Sechenov First Moscow State Medical University, 8 Trubetskaya str., 119991 Moscow, Russia

[4] Institute for Women's Health and Department of Mathematics, University College London, Gower Str., London WC1E 6BT, United Kingdom

[5] Neuroscience Research Institute, Samara State Medical University, 18 Gagarin Str., 443079 Samara, Russia

[6] Department of Mathematics, University of Leicester, University Rd, Leicester LE1 7RH, United Kingdom

*email: gordleeva@neuro.nnov.ru

## Abstract

Mammalian brains operate in a very special surrounding: to survive they have to react quickly and effectively to the pool of stimuli patterns previously recognized as danger. Many learning tasks often encountered by living organisms involve a specific set-up centered around a relatively small set of patterns presented in a particular environment. For example, at a party, people recognize friends immediately, without deep analysis, just by seeing a fragment of their clothes. This set-up with reduced "ontology" is referred to as a "situation". Situations are usually local in space and time. In this work, we propose that neuron-astrocyte networks provide a network topology that is effectively adapted to accommodate situation-based memory. In order to illustrate this, we numerically simulate and analyze a well-established model of a neuron-astrocyte network, which is subjected to stimuli conforming to the situation-driven environment. Three pools of stimuli patterns are considered: external patterns, patterns from the situation associative pool regularly presented to the network and learned by the network, and patterns already learned and remembered by astrocytes. Patterns from the external world are added to and removed from the associative pool. Then we show that astrocytes are structurally necessary for an effective function in such a learning and testing set-up. To demonstrate this we present a novel neuromorphic computational model for short-term memory implemented by a two-net spiking neural-astrocytic network. Our results show that such a system tested on synthesized data with selective astrocyte-induced modulation of neuronal activity provides an enhancement of retrieval quality in comparison to standard spiking neural networks trained via Hebbian plasticity only. We argue that the proposed set-up may offer a new way to analyze, model, and understand neuromorphic artificial intelligence systems.

## Introduction

The way the test data is organized, validated, as well as the method used to train learning systems can critically affect the result. Especially if the quality of learning is directly linked to survival. Mammalian brains are trained to survive, which is why they enable an animal to react quickly to patterns previously associated with dangerous situations. Hence, it is important to understand how such quick responses emerge in highly uncertain and complicated real-world operational conditions. We suggest that this ability is linked to the very specific mode of learning which can be called situation-based.

How many people did you meet yesterday? 5? 10? Was it difficult to recognize them? Psychologists say that our day-to-day activities impact our behavior. Recognition of patterns around us occurs in, what is called in psychology, a situation. Obviously, biological creatures that require less time to recognize a situation are getting an evolutionary advantage, they can escape a predator faster, get a higher chance of catching a prey, or when humans are concerned, earn more money. We all live in a situation-ridden world, and our life is based on recognition of patterns in current situation. But how is this recognition organized? We do not normally spend much time to recognize a friend, a fraction of a pattern is usually enough. Such quick processing of is provided by a very special structure of pattern learning. We learn patterns from a situation-based pool, and, since the number of patterns in the situation is limited, our pool usually is much smaller than in the whole external world. Such a situation-based structure of learning is visualized in the **Fig. 1**. There are three pools of patterns. Patterns from the huge external pool get into a situation-based pool, and then they become available for learning much more often than the ones arriving directly from the external pool. Hence, all patterns from a situation pool, except for the newcomers, are learnt and stored in the memory - the internal pool. Recognition is, hence, structure-associated, and patterns from the structure pool are recognized much easier and quicker patterns than from an external pool.

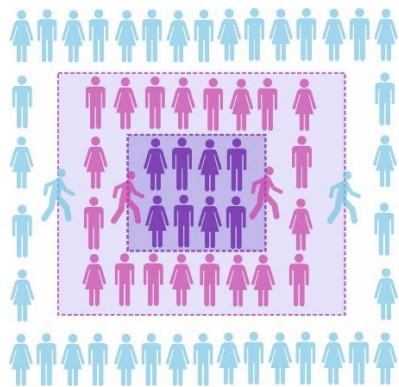

**Fig. 1 A diagram of situation-based model of data.** In this model, all data are partitioned into three pools of patterns. The largest circle of patterns (light blue) is the external world that contains a huge number of patterns. Situation-based pool of patterns is much smaller and includes only the patterns that we meet regularly within this situation (purple). Data patterns in the situation-based pool can be removed and replaced with patterns from the external world. Patterns from this pool are used for learning much more often than a random pattern from the external world pool, hence many of them are already learned, stored, and can be easily and quickly recognized by association with patterns stored in the system (dark blue lila).

Obviously, such situation-based structure has two main advantages: it is quick and requires less energy, which is very important in the biological world. Creatures adapted to such structure-oriented pattern of learning are more competitive and have an evolutionary advantage. This new data model is complementary to other important characteristics of learning and memory explored in the previous work, including high-dimensionality of the space of stimuli[1,2] and properties of data distributions conforming to the task of learning from few examples[3,4]. Importance of the problem was mentioned by[5,6].

All this leads to the key question whether there exists a structural organization of neural circuitry that is particularly suited for structure-based learning and that possesses characteristics of information processing in these circuits that are necessary to support this learning. In this paper, we propose relevant neural circuits that are particularly suited to facilitate situation-based learning. These circuits or networks combine conventional neurons and astrocytes.

The structural, metabolic, and homeostatic functions of astrocytes are well established[7]. Recently it has been revealed that astrocytes contribute to neural information processing via bidirectional exchange of regulatory signals with the neuronal elements. Astrocytes respond to neural activity by intracellular calcium elevations[8]. Calcium pulses in astrocytes induce the release of chemical transmitters (termed "gliotransmitters") which then regulate the synaptic gain of near and distant tripartite synapses at diverse timescales[9]. The data show that an astrocytes have an impact on local synaptic plasticity, neuronal network oscillations, memory and behavior[10,11,12]. Despite that the role played by astrocytes is not yet fully understood, these recent findings support the hypothesis that *cognitive processing and memory are not the result of neuronal activity only but of the coordinated activity of both astrocytes and neurons*[13]. Consequently, the most interesting research question is: whether a presence of astrocytes, which provide multiplex topology of a recognition network with different time and spatial scales of communication, facilitates the ability of the network to work with structure-associated learning? In this paper, we investigate this question and show numerically that neuron-astrocyte networks indeed play a key role in situation-based recognition. This function is also closely linked to the idea of local corrections in large neural networks working with big data[2].

**Related work**

Although astrocytic involvement in the information processing in the brain has been widely shown experimentally[10], there is a lack of computational studies of neural circuits focusing on astrocyte signaling in the context of learning and memory. The importance of computational modelling for developing better understanding of nature and findings answers to open questions is difficult to overestimate. Examples of works where such modelling brought new knowledge are numerous. In the area of astrocytes modelling, recent study[14] successfully demonstrated the self-repairing capability of distributed spiking neuron-astrocyte network in a robotic obstacle avoidance application. Nazari and colleages[15] studied the information transmission between the cortical spiking neural network and the cortical neuron–astrocyte network. They showed how cortical spiking network managed to improve its pattern recognition performance without the need for retraining by receiving an additional information from neuron-astrocyte network. In addition, scholars proposed several digital implementations of astrocytic dynamics[16] and neuron-astrocyte interaction[17,18]. In our previous works, we investigated how the astrocyte-induced dynamic coordination in the neuronal ensembles[19,20] induces the generation of integrated information sets[21,22,23]. Moreover, we showed that biologically-inspired spiking neuron-astrocyte network can implement the multi-item short-term memory[24,25]. We revealed that several information patterns can be maintained in memory at the time scale of calcium elevation in astrocytes, while the readout by the neurons due to the astrocyte-induced activity-dependent short-term synaptic plasticity resulted in local spatial synchronization in neuronal ensembles. We further showed that spiking neuron-astrocyte network can reliably store not only binary but also analogous information patterns in short-term memory[26]. However, the work in this paper goes much further and proposes a new bio-inspired two-net spiking neuron-astrocyte network (SNAN) for more complex learning tasks, in which SNAN is implemented for associated learning.

**Significance**

In this paper, we present three key findings: (1) a novel approach to formalizing machine learning data, namely, the temporal organization of the data as opposed to the widely accepted IID data sampling; (2) a novel neuromorphic computational model for short-term memory implemented by SNAN; and, (3) a proof, through rigorous computational experiments, that SNAN tested on synthesized data with selective astrocyte-induced modulation of neuronal activity may provide an enhancement of retrieval quality in comparison to a standard SNN trained via Hebbian plasticity. The proposed SNAN is a hybrid system, which combines the fast-spiking neural networks pre-trained by the Spike Timing Dependent Plasticity (STDP) rule with the general data set, and a slow astrocytic network, which provides time-dependent data buffering via calcium activity and gliatransmitter-induced spatial-temporal coordination of neural network activity.

**Situation-based learning in spiking neuron-astrocyte network model**

The concept of the proposed situation-based memory model is schematically summarized in **Fig. 2**. A new biologically motivated computational model of short-term memory is implemented through interaction of neural and astrocytic networks. The model acts at multiple timescales: at a millisecond scale of firing neurons and the second scale of calcium dynamics in astrocytes. The neuronal network consists of randomly sparsely connected excitatory and inhibitory spiking neurons with plastic synapses. To train synapses in neural network, we used the traditional spike-timing-dependent plasticity (STDP) rule. Astrocytes track the neural activity and respond to it by intracellular calcium elevations, which trigger the release of gliotransmitters. Gliotransmitter-induced short-term synaptic plasticity results in local spatial synchronization in neuronal ensembles. The short-term memory realized by such astrocytic modulation is characterized by one-shot learning and is maintained for seconds. The astrocytic influence on the synaptic connections during the elevation of calcium concentration implements Hebbian-like synaptic plasticity differentiating between specific and non-specific activations. Composed of two building blocks, e.g., fast-spiking neurons and slow astrocytes, the proposed memory architecture eventually demonstrated synergetic functionality in loading information. The readout of this memory by the neuronal block and storage implemented by the astrocytes.

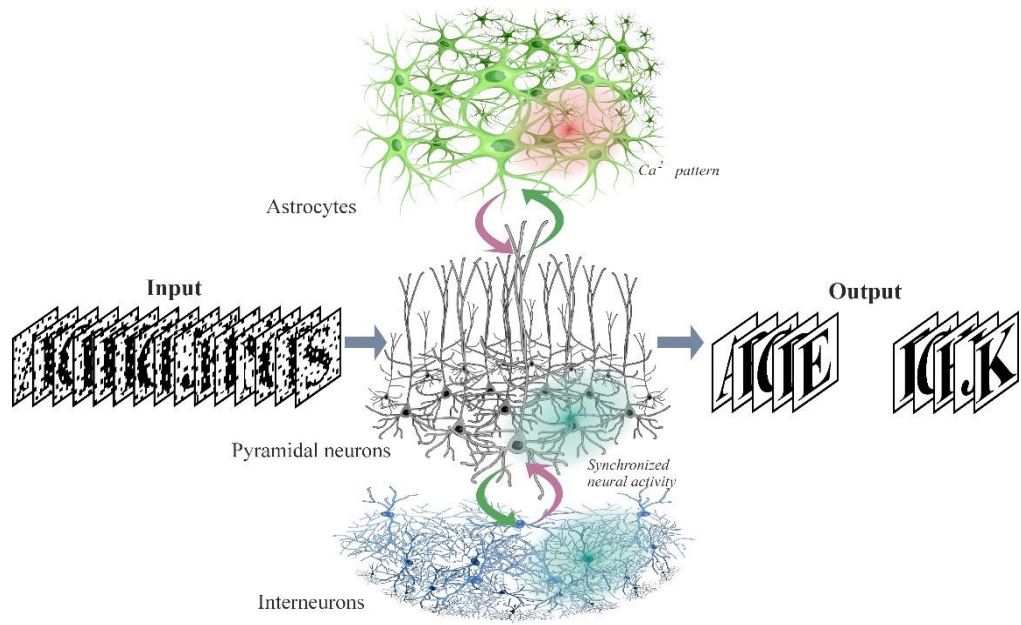

**Fig. 2 Concept of situation-based memory operation in the spiking neuron-astrocyte network model.**

**Spiking neuron-astrocyte network architecture.** The architecture of the proposed SNAN is illustrated in **Fig. 3**. The SNAN includes three interacting layers: the layer of pyramidal neurons, the layer of interneurons, and the astrocytic layer. An input signal encoded as two-dimensional patterns was applied to the first layer. The first layer consists of 6241 (79×79) synaptically coupled pyramidal neurons, which are connected randomly with their connection length determined by the exponential distribution. To maintain the balance of excitation and inhibition during neuronal activity, the layers of pyramidal neurons and interneurons communicate bidirectionally. Astrocytes generating calcium signals are connected by local gap junction diffusive couplings. To design the neuronal and astrocytic layers interaction, we followed the approach proposed in our previous works[27,28]. Calcium elevations occur in response to the increased concentration of the neurotransmitter released by pyramidal neurons when a group of them fire coherently. In turn, gliotransmitters are released by activated astrocytes modulating the strength of the synaptic connections in the corresponding neuronal group. The output signal is taken from frequencies of transient discharges of pyramidal neurons.

A detailed information concerning the models and the description of parameters is provided in the **Methods**.

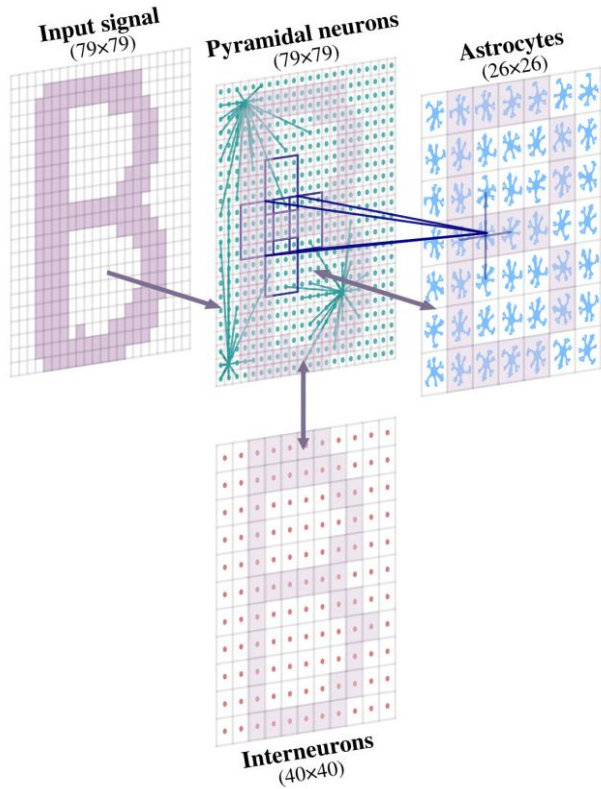

**Fig. 3 A spiking neuron-astrocyte network topology.** The SNAN includes three interacting layers: the layer of pyramidal neurons, the layer of interneurons, and the astrocytic layer. First layer (79×79) consists of synaptically coupled pyramidal neurons. The pyramidal neurons bidirectionally communicate with the interneurons from the second layer (40×40). Ratio of pyramidal neurons to interneurons in the model is chosen in accordance with the experimental observations and computational model of the cortex[29], where 80% of the CSN neurons are pyramidal neurons and 20% are interneurons. Astrocytes are connected by a local gap junction diffusive couplings and represent a two-dimensional square lattice with a dimension 26×26. We focus on the bidirectional interaction between the first neuronal and astrocytic layers. Each astrocyte is interconnected with an ensemble of $N_a$=16 pyramidal neurons with dimensions 4×4 (red lines) with overlapping in one row and one column. An input signal encoded as a two-dimensional pattern is applied to the first layer.

**Training and Testing Protocol.** To train and test the proposed SNAN, we use the alpha-digits data set[30] which consists of $P$ binary images of digits and capital letters of size $W \times H$ pixels. The input patterns are fed to the layer of pyramidal neurons. Each image pixel corresponds to a neuron, which receives a rectangular excitatory pulse, $I_{app,i}$, with length $t_{stim}$ and amplitude $A_{stim}$ for training (with $t_{test}$ and $A_{test}$ in case of testing). On average there are 950 neurons under stimulation (15% of the network) in a training image. Training samples were presented to highly overlapped neuronal populations (an average for 40 training samples overlapping was 51%). The output signal was read out according to the firing rates of pyramidal neurons.

**SNN pre-training.** First, we pre-trained the synaptic connections only in the spiking neuronal network consisting of pyramidal neurons and interneurons without taking into account the influence of astrocytes. During pre-training, each of $P$ patterns was presented to the neuronal network 10 times in random order. After the pre-training was completed, the network weights were fixed. To test the training quality, we calculated the correlation of recalled patterns with the ideal samples according to the procedure described in **Methods**. In the cued recall, we applied a shorter inputs with lower amplitude ($t_{test}$, $A_{test}$) to the network. These inputs were spatially distorted by high-level random noise matching the training samples.

**Situation-based learning in SNAN.** To implement the situation-based learning in the proposed SNAN, we use the following protocol. After the SNN pre-training, we turn on the bidirectional interaction between pyramidal neuron layer and astrocytic layer. To let the astrocytic network generate the first calcium pattern, we apply the initial pool of patterns to SNAN. This pool consists of 7 (seven) randomly selected patterns from the general data set used in the SNN pre-training. Each pattern was presented 10 (ten) times with the addition of a random 5% "salt and pepper" noise. After a break (approximately 650 ms) necessary for the formation of calcium impulses in pattern-associated astrocytes, we start the ongoing training-testing process of the SNAN in real time. This situation-based learning process can be conventionally divided into a sequence of cycles, which follow each other continuously.

Every test cycle starts with training of the SNAN on one new pattern which was absent in the initial pool and was randomly chosen from the general data set. After that, we test the storage of all patterns from the initial pool in memory. Throughout the article, we use the term 'memory' referring to the ability of pattern recall in presence of perturbations. We present the SNAN with the 7 test patterns which match the patterns from the initial pool, but have a shorter lengths, lower amplitude ($t_{test}$, $A_{test}$) and which are spatially distorted by high level (20%) random noise. To identify the memory performance, we analyze the quality of the recalled patterns. In the next cycle, one pattern from the initial pool is replaced by a new pattern which has been learned in the previous cycle. Thus, after 7 cycles all patterns from the initial pool are substituted by new patterns from the general pre-training data set. This procedure can be performed endlessly allowing the system to work with all patterns from the general data set in a situation-based mode. **Fig. 4** illustrates the time scheme of the training and testing protocol. The values of the stimulation parameters are listed in **Table 4**.

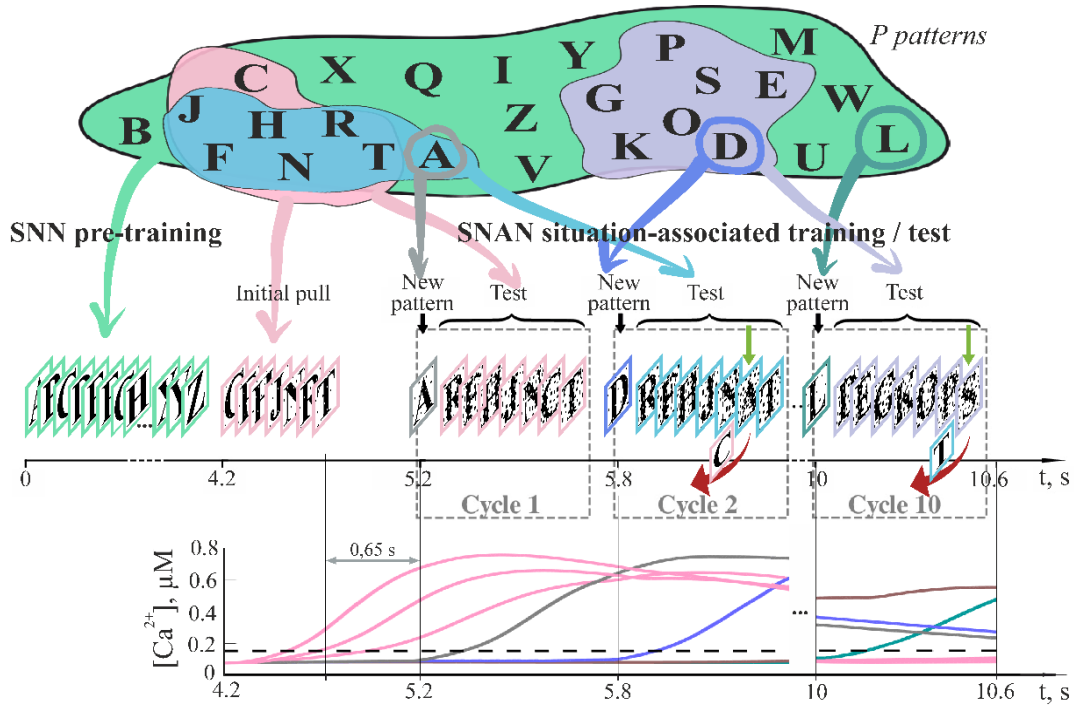

**Fig. 4 Training and testing protocol.** For training and testing of the proposed SNAN we use the alpha-digits data set consisting of $P$ binary images of digits and capital letters. The input patterns are fed to the layer of pyramidal neurons. First, we pre-train the spiking neuronal network consisting of pyramidal neurons and interneurons without taking into account the influence of astrocytes. During pre-training, each $P$ pattern is presented to the neuronal network 10 times in random order (green color). After the pre-training is completed, the synaptic weights are fixed. To implement the situation-based learning in the proposed SNAN, we use the following protocol. After the SNN pre-training, we turn on the bidirectional interaction between pyramidal neurons and astrocytic layers. To let the astrocytic network generate the first calcium pattern, we apply the initial pool of patterns to SNAN. This pool consists of 7 (seven) randomly selected patterns (pink color) from the general data set used in the SNN pre-training. Each pattern is presented 10 times with the addition of a random 5% "salt and pepper" noise. After a break (approximately 650 ms) necessary for the formation of calcium elevations in the pattern-specific astrocytes (examples of astrocytic $Ca^{2+}$ signals are shown in colors corresponding to the patterns), we start ongoing training-testing process of the SNAN in real time. This situation-based learning process can be conventionally divided into a sequence of cycles, which follow each other continuously. Every test cycle starts with training of the SNAN on one new pattern (e.g. pattern "A", grey color), which was absent in the initial pool and was randomly chosen from the general data

set. After that, we test the memorization of all patterns from the initial pool - "Cycle 1". We present the SNAN with test patterns which have been spatially distorted by high-level noise. To identify the memory performance, we analyze the quality of the recalled patterns. In the next cycle, "Cycle 2", one pattern from the initial pool (pattern "C") is replaced by new pattern which has been learned in the previous cycle (pattern "A"), this models situation-based environment. Thus, after *N* cycles all patterns from the initial pool are substituted by new patterns from the general pre-training data set. This procedure can be performed endlessly allowing the system to work with all patterns from the general data set in situation-based mode.

## Results

**SNN memory performance.** First, we determine the size of the general data set that can be loaded in memory of the SNN and used for implementation of the situation-based learning in the proposed SNAN. For this, we pre-train the SNN on the data set of different sizes and test the quality of memory maintenance of the SNN in cued recall. Information retrieval is organized by the application of a cue sample representing one pattern from the memory set distorted by "salt and pepper" noise. The dependencies of the correlation between the SNN cued recalls and the ideal target samples (averaged over all test patterns ± standard deviation) on the data set size are shown in **Fig. 5** by red curves. Two cases were considered for test images distorted by 20% (**Fig. 5A**) and 30% (**Fig. 5B**) noise levels. The maximum correlations between SNN recalls and non-target sample averaged over all test patterns and the maximum correlation between target and non-target samples are presented by blue and green curves in **Fig. 5**, respectively. According to the results obtained, the considered SNN can learn up to 40 patterns. In further analysis, we used data set sizes of 20 and 40 patterns for comparison.

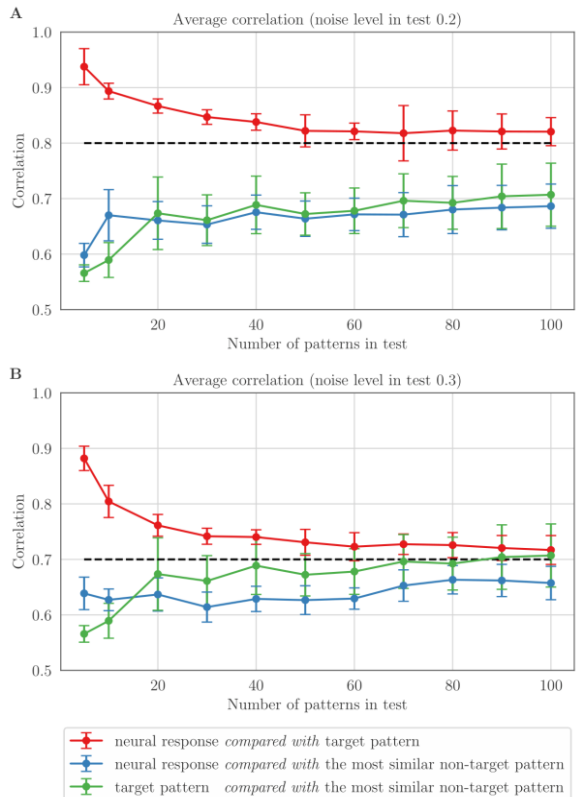

**Fig. 5 The pre-trained SNN's memory performance.** The correlations between SNN recalls and the ideal target samples for different data set size are marked by red. The maximum correlations between SNN recalls and non-target sample are marked by blue. The maximum correlation between target and non-target samples are marked by green. Average means over all test patterns ± standard deviation are shown for 20% (A) and 30% (B) noise levels in test images. The dotted line indicates test patterns correlation.

**SNAN situation-based learning performance.**

**Astrocytic contribution to the SNAN memory performance.** To assess the contribution of astrocytes in information processing and memory formation in neuron-astrocyte networks, the pre-trained SNN was bidirectionally connected to the astrocytic layer. To start the process of the SNAN situation-based learning, we load the initial pool of 7 patterns to the system by applying the inputs (**Fig. 6A, D, G**) to the pyramidal neuronal layer. The activity of pattern-specific neuronal subnetworks (**Fig. 6B, E, H**) induces the generation of calcium signals in corresponding astrocytes. Due to the fact that calcium dynamics in astrocytes has slow scale, the overlapped spatial calcium patterns in astrocytic layer for different samples coexist for several seconds (**Fig. 6C, F, I**).

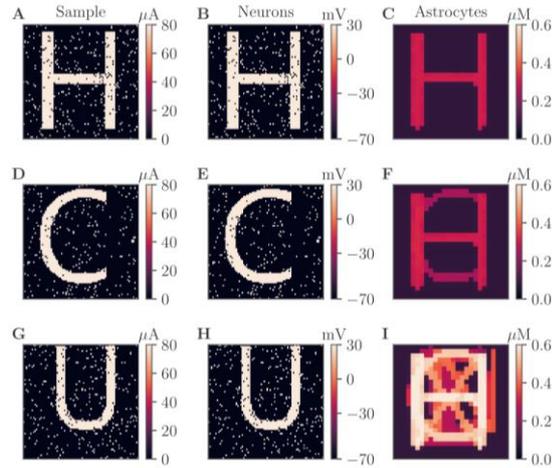

**Fig. 6 The example of SNAN training on the patterns from initial pool (Fig. 4).** (A, D, G) The first, second, and the 7th training patterns from initial pool, respectively. (B, E, H) Responses of the pyramidal neuronal layer to the patterns. The values of the membrane potentials are shown. (C, F, I) Intracellular $Ca^{2+}$ concentrations in the astrocytic layer.

Then we ran the ongoing process of situation-based learning according to the approach described in the Section "**Situation-based learning in SNAN**" and illustrated in **Fig. 4**. Briefly, in each of the 10 cycles, we loaded a new pattern from pre-training data set to the SNAN and tested the patterns memorized in the previous cycles. A constant number of patterns in the cycle was maintained by deleting one randomly selected pattern during each cycle. Test patterns were applied to the pyramidal neurons with 20% level noise, and the SNANs cued recalls in the values of the neuronal firing rate were read out. Examples of input test images from several cycles and the systems retrievals are shown in **Fig. 7**.

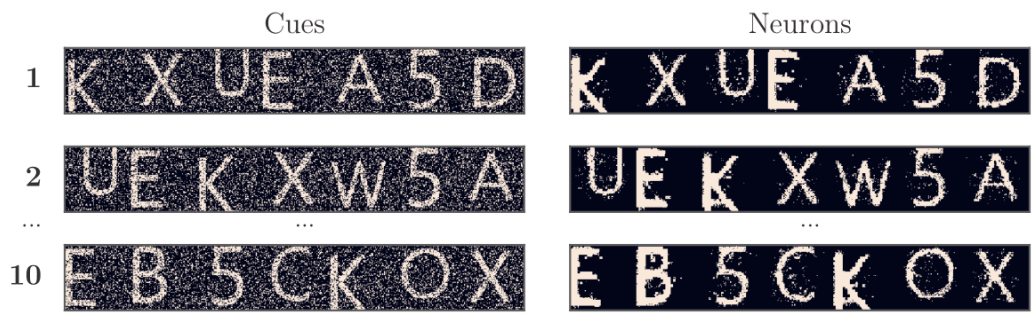

**Fig. 7 The example of the SNAN test patterns from the general pre-trained data set (Fig. 4).** Three testing cycles are shown. Left panel presents the testing images with 20% "salt-and-pepper" noise. Right panel presents the cued recalls in the pyramidal neuronal layer. The averaged firing rate on the test time interval for each neuron is shown.

To estimate the astrocytic impact on the memory formation in the SNAN, we calculated the dependencies of recall correlation with samples on the noise level. First the test was run with astrocytic modulation of synaptic transmission in the SNN and then without it (**Fig. 8**). The test involved 20 and 40 patterns from the pre-trained data set. The differences in correlation between the recalled pattern and noisy input clearly show that astrocytes steadily improve the quality of the system retrieval up to 10% for high noise levels (red curve in comparison with blue curve). The reason for such recalls enhancement is that a short presentation of the cue to the neural network evokes the additional astrocytic-induced spike in the synaptic strength between stimulus-specific neurons, which results in a local spatial synchronization in the whole stimulus-specific neuronal population.

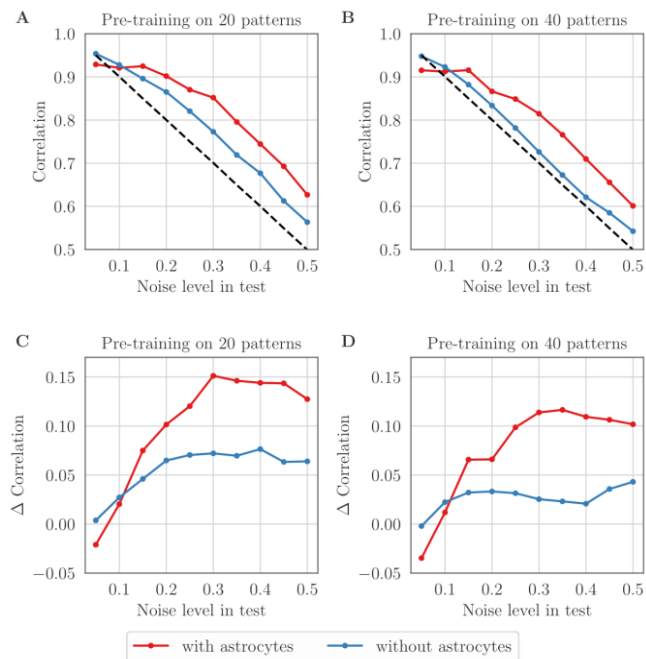

**Fig. 8 The astrocyte-induced enhancement of the memory performance in the proposed SNAN.** (A, B) The correlations between SNAN recalls and the ideal samples dependent on noise level in testing patterns with astrocytic modulation of synaptic transmission in neural network (red curves) and without it (blue curves). (C, D) The difference between correlations of systems recalls and test patterns. (A,C) and (B, D) correspond to data set sizes of 20 and 40 patterns, respectively. The dotted line indicates test patterns correlation.

**Contribution of the SNN pre-training to the SNAN memory performance.** Next, we evaluate the contribution of the neural network learning to the SNAN memory performance according to synaptic weights adjustment via the STDP rule. For this, we compare the memory performance of the three

SNAN types (i) with synaptic connections trained according to the STDP rule, (ii) with randomly mixed synaptic weights after the SNN pre-training, and (iii) with fixed synaptic weights without the SNN pre-training. **Fig. 9A** shows the changes in the correlation of the SNANs cued retrievals relative to the input noise patterns for these cases with and without astrocytic influence on neural activity. The best levels of recall correlations were demonstrated by the proposed SNAN trained by the STDP rule with astrocytic modulation of synaptic transmission, followed by the SNAN with mixed synaptic weights and astrocytic modulation, and then the pre-trained SNN without astrocytes. The worst results were shown by networks without astrocytic modulation of synaptic transmission and without training of synaptic connections. Interestingly, astrocyte-induced enhancement of synaptic transmission in the sample-specific neuronal subnetworks can provide good quality retrieval in the system even for neural networks with mixed weights of synaptic connections (blue line in **Fig. 9A**).

**Effect of the synaptic connectivity strength on the SNAN memory performance.** Next, we studied the influence of synaptic connectivity architecture in the neural layers of the SNAN on the correlation of the system recalls. We specifically focused on the weight of synaptic connections between layers and inside the pyramidal neuronal layer. Higher inhibition of the system (**Fig. 9B**) due to the increase of the maximum synaptic weights of the connection from interneurons layer to pyramidal neurons, $w_{synIEmax}$, induces the SNN memory performance decline (red dashed line), but does not affect the SNAN memory performance (red line). This can be explained by the fact that samples in training were applied to highly overlapped neuronal population. Such subnetworks of interneurons corresponding to several patterns provide strong inhibition of the signal propagation in sample-specific population of pyramidal neurons and prevent correct recall. However, this can be compensated by the stimul-specific astrocyte-induced enhancement of excitatory synaptic transmission. On the contrary, increasing the maximum excitatory synaptic strengths in the pyramidal neuronal layer, $w_{synEEmax}$, results in astrocyte-induced overactivation of the SNAN and a decrease in the recall quality (**Fig. 9C** red and green lines).

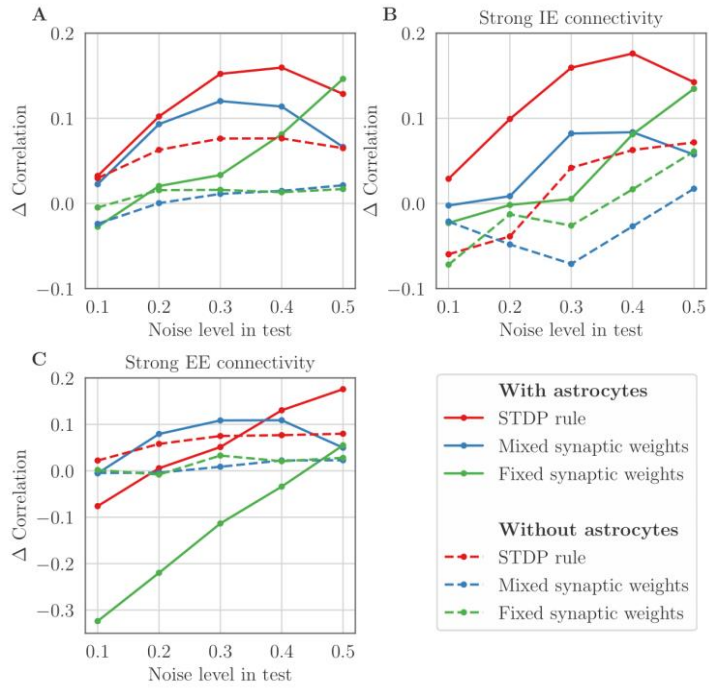

**Fig. 9 The difference between correlations of the SNANs recalls and test patterns.** The memory performance was shown for three SNAN types with and without astrocytic influence: (i) with synaptic connections trained according to the STDP rule, (ii) with randomly mixed synaptic weights after the SNN pre-training, and (iii) with fixed synaptic weights without the SNN pre-training. (A) $w_{synIEmax}$=0.05, $w_{synEEmax}$=0.05; (B) Strong connections from interneurons to pyramidal neurons $w_{synIEmax}$ =0.15, $w_{synEEmax}$ =0.05; (C) Strong connections inside the pyramidal neurons layer $w_{synIEmax}$ =0.05, $w_{synEEmax}$ =0.07.

**Capacity of the situation-based memory in the SNAN.** The situation-based memory capacity in the proposed SNAN is determined by the duration of $Ca^{2+}$ signals in astrocytes. Duration of astrocytic $Ca^{2+}$ elevations is determined by the intrinsic mechanisms of the $IP_3$-evoked calcium release from the endoplasmic reticulum in astrocytes, which is described by the biophysical model[31] used in this study. Brief application of the cue samples during testing results in prolongation of $Ca^{2+}$ elevations in astrocytes and, thus, in the increased storage time of patterns in the memory of the SNAN. On average, the $Ca^{2+}$ signals duration in astrocytes is 3.8 s, which can support the situation-based learning during 9 cycles on 15 different patterns.

**Impact of the overlapping level in samples on the SNAN memory performance.** The pre-trained spiking neural network can retrieve the correct samples from test images distorted with 20% noise level with the average correlation level of 96%. This, however, applies only to non-overlapping patterns without additional effect of astrocytic modulation, since even a small sample overlapping results in chimeras generation in the solely neuronal network model. To characterize the impact of the overlapping level in training samples on the SNAN memory performance, we use rectangles of different sizes displaced at the fixed number of pixels relative to the neighbor as information patterns (**Fig. 10A**). In this case, in contrast to the used alpha-digit data set, the level of overlapping between the neighboring patterns can be precisely specified.

After the SNN pre-training on 40 patterns with fixed overlapping, we use the situation-based training and testing protocol for the SNAN described above with little modifications. To be sure that sample overlapping level inside one cycle is constant between all patterns, we apply samples to the SNAN sequentially (not in random order as before). The example of corresponding calcium activity in astrocytic layer is shown in **Fig. 10B**. The dependencies of correlation level of the SNANs cued recalls on different overlapping levels of samples are shown in **Fig. 11** for SNAN with astrocytic modulation of synaptic transmission and without it. Results show that including the astrocytic modulation of synaptic transmission into spiking neural network with connections trained according to the Hebbian plasticity leads to a robust improvement of the system retrieval performance for almost all levels of sample overlapping, excluding the highest levels ($>80\%$). It is important to note that the contribution of astrocytes is especially significant for a high noise value in cue samples (comparing **Fig. 11A** with **Fig. 11B**). On average, in range of samples with overlapping level from 0 to 0.9, the astrocyte-induced enhancements of retrieval quality (in particular, the correlation of cued recalls with ideal samples) amounts to 5% for test samples distorted by 20% noise and 20% for - 30% noise in cue samples. **Fig. 11** shows that even for huge pattern overlapping, spurious correlations never dominate in that sense as accuracy of our system is always equal to 100%. Correlation with the target sample always exceeds correlation with the wrong sample.

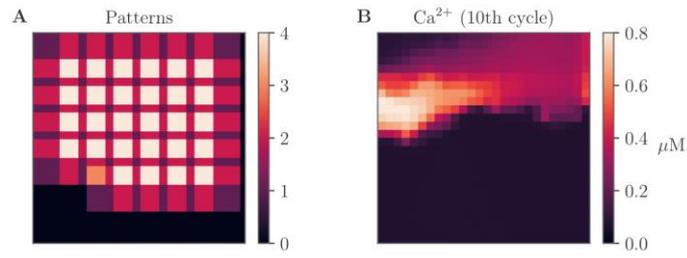

**Fig. 10 (A) The example of data set used for evaluation of the impact of the samples overlapping on the SNAN memory performance.** The figure shows the case for samples of size 17×17 pixels with overlapping in 7 pixels which corresponds to overlapping in 41.18% between the neighboring patterns. **(B) The corresponding calcium activity in astrocytic layer.**

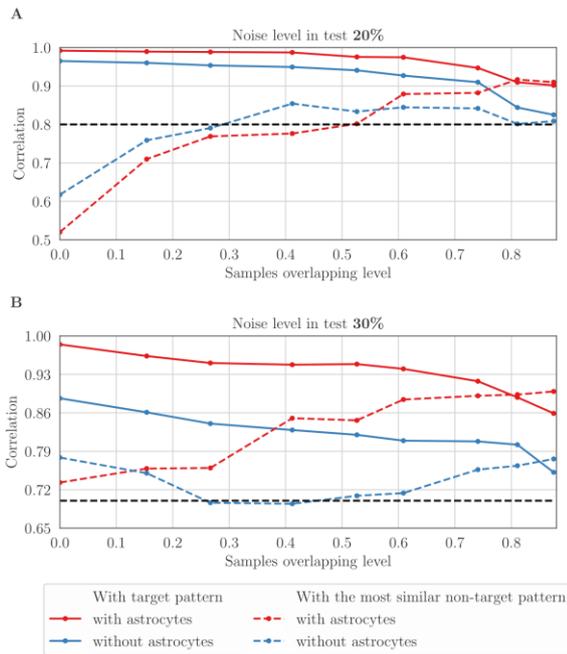

**Fig. 11 Impact of the overlapping level in samples on the SNAN memory performance.** The dependencies of correlation level of the SNANs cued recalls and samples for different level of sample overlapping are shown for SNAN with astrocytic modulation of synaptic transmission and without it for 20% (A) and 30% (B) noise level. Blue and red dotted lines show correlation of system recalls with the most similar non-target samples. Black dotted lines indicate test patterns correlation.

## Conclusion

This paper presents a novel approach to temporal non-IID data organization for machine learning in spiking neuronal networks. The effectiveness of data formalization in situation-based pools is demonstrated by the short-term memory task implemented by the brain-inspired spiking neuron-astrocytic network (SNAN). The SNAN includes a layer of principle (pyramidal) neurons supplied by a group of inhibitory interneurons. Synaptic connections in the pyramidal layer self-adjust adaptively according to the Hebbian-like spike timing dependent plasticity (STDP). Following morphological brain synaptic organization, the pyramidal neurons are accompanied by astrocytes organized in the form of a layer network (see **Fig. 3**). Astrocytic modulation of neuronal activity represents the activity-dependent short-term synaptic plasticity, which induces the stimulus-specific local spatial synchronization in neuronal ensembles. The synergistic interplay between fast spiking neuronal network trained on the general data set and slow astrocytic syncytia provides buffering of situation-based data pools by the selective coordination of neuronal signaling, which results in successful storage and retrieval of highly overlapped information patterns. We demonstrated that the astrocyte-induced influence on synaptic transmission results in 10% enhancement of spiking neural network memory performance in terms of correlation level between the cued retrievals and samples for strong 50% overlapped patterns.

## Discussion

The results obtained in the paper could be instrumental for the development of the brain-like (e.g. "strong") artificial intelligence. Inspired by the brain structural and functional organizational hierarchy, neuromorphic hardware systems that implement spike-driven computations could potentially be capable of implementing energy-efficient machine intelligence[32]. In addition, the possibility to enhance learning performance by astrocytes is an important milestone in the ongoing discussion of the role astrocyte-neuronal network interactions in brain processing[10]. Specifically, we have investigated functional roles of different players, e.g. neurons, synapses, plasticity, astrocytes, in the implementation of cognitive information processing tasks in the brain. In particular, it was interesting to observe how the interplay of synaptic changes by STDP and by the gliotransmitter modulations improve memory performance (see, for example, **Fig. 9**).

The STDP-based plasticity represents a key biophysical mechanism of learning in spiking neuronal networks. In memory tasks, the synaptic weights are adjusted following a training protocol by sequential image application. Indeed, we also verified that STDP provided successful in training and information retrieval with certain degree of fidelity. In this context, the interneurons balanced network firing by depression and, hence, we can safely assume that they are responsible for lateral inhibition via "selecting" stimulus specific excitation routes. This solely neuronal story cannot resolve the problem of "overlapping populations" when different input patterns stimulate similar neuronal groups (up to 50% of overlaps in our samples). Obviously, the synaptic plasticity alone cannot resolve this problem, as it will inevitably lead to false recalls and the decreased performance. However, our research shows that astrocytes can significantly improve this situation.

The astrocytic calcium operates at much slower time scales, hence, the astrocytes cannot significantly affect the fast dynamics of neurons and synapses at the time scale of single image processing. Moreover, the calcium excitability has a gradual character[33]. It provides a proportional response to stimuli with different intensities. Thus, the stronger the activation of pyramidal neurons in terms of their discharges intensities over interval of dozen of seconds, the higher the calcium response in the corresponding astrocytes. This way the astrocytes corresponding to the overlapping areas generate larger signals. In turn, they send back different level of modulations during the recall processing. Furthermore, patterns with a high degree of overlap can be successfully resolved, which gives a noticeable increase in the retrieval fidelity.

At functional level, astrocytes supplement neuronal processing by an amplitude modulation in addition to rate encoding by all-or-none firing neurons. At the same time, being distributed in time the astrocytic modulation provides a dynamic separation of overlapping patterns. It is very similar to the reservoir computing in machine learning with traditional artificial neurons[34,35]. Here, the atrocytes serve as a reservoir naturally "predicting" correct retrieval due to dozen of seconds of stored history.

In general, decoding the physiological meaning of the spatial-temporal $Ca^{2+}$ signalling in astrocytes, its computational properties, and impact on neuronal signalling remains a major challenge in modern neurobiology[8]. Integration of astrocytic signaling in cognitive processing has implications for understanding the basis of cognitive dysfunction and development of new therapeutic strategies[10,36,37]. To provide a stronger link of the proposed SNAN with neuroscience, next it might be interesting to employ the mechanisms of intracellular integration of $Ca^{2+}$ signals in astrocytes[25].

## Methods

In this section, the SNAN architecture is described in detail together with the STDP learning rule and neuron/astrocytic models. Specifically, we start with biological realistic models of neuronal, astrocytic networks that capture the essence of the biological interplay between these cells, at the same time minimizing the computational overhead. Then we describe a communication between pyramidal neurons and astrocytes at tripartite synapses.

**Neural network.** Among the many existing biological plausible spiking neuron models[38,39,40,41], we have chosen simplified Izhikevich model[42] as computationally efficient for modeling networks. The dynamics of neuronal membrane potential is given by[42]:

$$\frac{dV_i}{dt} = 0.04V_i^2 + 5V_i - U_i + 140 + I_{app,i} + I_{syn,i};$$
$$\frac{dU_i}{dt} = a(bV_i - U_i);$$
(1)

with the auxiliary after-spike resetting

$$if\ V_i \geq 30\ mV, then \begin{cases} V_i \leftarrow c \\ U_i \leftarrow U_i + d \end{cases}$$
(2)

where the subscript $i$ corresponds to the neural index, $V_i$ is the neuronal membrane potential in mV, time $t$ in ms. The applied current $I_{app,i}$ simulates the input signal, $I_{syn,i}$ is the synaptic current. The parameter descriptions and their values used in this work can be found in **Table 1** in **Methods**.

The total synaptic current injected from all synapses of $i^{th}$ neuron is described by[43,44]:

$$I_{syn,i} = \sum_{k=1}^{N_i} \frac{w_{syn,k}(E_{syn} - V_i)}{1 + \exp(-V_{pre,k}/k_{syn})},$$
(3)

where $N_i$ is the total number of synapses, $w_{syn,k}$ is the weight of the $k^{th}$ synapse associated with neuron, $V_{pre}$ is the membrane potential of the presynaptic neuron, $E_{syn}$ is the synaptic reversal potential. $E_{syn}$ = -90 mV for the inhibitory synapse and $E_{syn}$ = 0 mV for the excitatory. Parameter $k_{syn}$ denotes the slope of synaptic activation function threshold. We neglect the synaptic and axonal delays in system for simplicity.

Pyramidal neurons interact with each other (connection type: EE) and with interneurons (EI). Interneurons communicate with pyramidal neurons (IE) and are not interconnected. The architecture of synaptic connections between neurons is non-specific (random) with different parameters within excitatory and inhibitory layers, as well as between layers, which is described further below. A detailed list of parameters values of synaptic connection organization can be found in **Table 1**. The number of output connections per each neuron is fixed at $N_{out}$. Each postsynaptic neuron is randomly selected in polar coordinates. The distances between neurons $r$ are determined by the exponential distribution $f_R(r)$, and the angles $\varphi$ are chosen from a uniform distribution in the range [0; 2π]:

$$f_R(r) = \begin{cases} \frac{1}{\lambda}\exp(-r/\lambda) \geq 0, \\ 0, r < 0. \end{cases} \quad (4)$$

Taking into account the difference in the sizes of the layers, the coordinates of postsynaptic neurons are calculated as follows:

$$EE: x_{post} = [x_{pre} + r\cos(\varphi)], y_{post} = [y_{pre} + r\sin(\varphi)];$$

$$EI: x_{post} = [K_1^{-1} x_{pre} + r\cos(\varphi)], y_{post} = [K_2^{-1} y_{pre} + r\sin(\varphi)]; \quad (5)$$

$$IE: x_{post} = [K_1 x_{pre} + r\cos(\varphi)], y_{post} = [K_2 y_{pre} + r\sin(\varphi)],$$

where $x_{pre}$, $y_{pre}$ denote the coordinates of the presynaptic neuron, $x_{post}$, $y_{post}$ are the coordinates of the postsynaptic neurons, $K_1 = W/W_1$, $K_2 = H/H_1$. Coordinates are picked repeatedly in case of duplicated connection (random selection was a process without replacement).

In the proposed SNAN, the synaptic weights dynamically adjust during training only for EE and IE types of synaptic connections. The synaptic weights for EI synapses are fixed and equal to $w_{synEI}$ = 0.1. The initial weights of the synapses between pyramidal neurons (EE) and interneuron–pyramidal neuron (IE) are $10^{-4}$. The maximum weights are limited to values $w_{synEEmax}$, $w_{synIEmax}$. The STDP rule updates the synaptic weights according to the timing difference between the pre and postsynaptic spikes, and is described by:

$$\delta w_{synEE,k}(\Delta t) = \begin{cases} g_{synEE} \exp\left(\frac{\Delta t}{\tau}\right), \Delta t \leq 0, \\ -g_{synEE} \exp\left(\frac{\Delta t}{\tau}\right), \Delta t > 0; \end{cases} \quad (6)$$

$$w_{synEE,k} \in [10^{-4}, w_{synEEmax}],$$

where $\delta w_{synEE,k}(\Delta t)$ is used to update the synaptic weight, $\Delta t$ is the time difference between presynaptic and postsynaptic spikes, $g_{synEE}$ is the plasticity window height, $\tau$ control the width of the plasticity window, and they are 20 ms in our model. Training of synaptic connections from interneurons to pyramidal neurons is organized so that interneurons activated by pyramidal neurons inhibit all subnetwork of pyramidal neurons that were not active during the presentation of the training pattern. In such way, the weights of IE synapses are updated according the following:

$$\delta w_{synIE,k}(\Delta t) = \begin{cases} g_{synIE} \exp\left(\frac{\Delta t}{\tau}\right) H(f^* - f), \Delta t \leq 0, \\ -g_{synIE} \exp\left(\frac{\Delta t}{\tau}\right), \Delta t > 0; \end{cases} \quad (7)$$

$$w_{synEE,k} \in [10^{-4}, w_{synIEmax}],$$

where $\Delta t$ is the time difference between presynaptic and postsynaptic spikes, $g_{synIE}$ is the plasticity window height, $\tau$ control the width of the plasticity window, and they are 20 ms in our model. $f$ and $f^*$ are the actual firing rate (i.e. a running average over 10 ms) and threshold firing rate of the postsynaptic pyramidal neuron, respectively. $H$ is the Heaviside step function.

**Astrocytic network.** The astrocytic layer consists of 676 cortical astrocytes connected with only nearest neighbors. It has been experimentally shown that an individual cortical astrocyte contacts on several neuronal somatas and hundreds neuronal dendrites with some overlapping in the spatial territories corresponding to different astrocytes in the cortex[45]. Such an organization of neuron-astrocyte interaction allows the astrocytes to integrate and coordinate a unique volume of synaptic activity. Following experimental evidences, each astrocyte in the SNAN bidirectionally interacts with ensemble of $N_{AE}=16$ pyramidal neurons with some overlapping. Spiking neuronal activity induces the release of neurotransmitter (glutamate) from the presynaptic terminals into the synaptic gap. The released glutamate binds to the metabotropic glutamate receptors (mGluRs) on the astrocyte membrane and triggers the production of inositol 1,4,5-trisphosphate (IP3) in astrocytes, which is followed by the generation of a calcium pulse. The Ullah model[31] is used to describe the dynamics of the intracellular concentrations of IP3 and $Ca^{2+}$ in astrocytes:

$$\frac{d[Ca^{2+}]_m}{dt} = J_{ER} - J_{pump} + J_{leak} + J_{in} - J_{out} + J_{Gca};$$

$$\frac{dh_m}{dt} = a_2 \left( d_2 \frac{[IP_3]_m + d_1}{[IP_3]_m + d_3} (1 - h_m) - [Ca^{2+}]_m h_m \right); \qquad (8)$$

$$\frac{d[IP_3]_m}{dt} = \frac{[IP_3]^* - [IP_3]_m}{\tau_{IP3}} + J_{PLC\delta} + J_{glu} + J_{Gip3},$$

where $m$ ($m = 1, \ldots, 676$) is the astrocyte index. $[Ca^{2+}]$, $[IP_3]$, $h$ are the cytosolic calcium and IP3 concentrations and fraction of activated IP3 receptor on the endoplasmic reticulum (ER) membrane, respectively. $J_{ER}$ is $Ca^{2+}$ flux from the ER to the cytosol, $J_{pump}$ is the pump flux from cytosol to ER, $J_{leak}$ is the leakage flux from the ER to the cytosol. The fluxes $J_{in}$ and $J_{out}$ describe the exchange of calcium with the extracellular space. $J_{PLC\delta}$ describes the production of IP3 by phospholipase C$\delta$ (PLC$\delta$), $J_{glu}$ describes the glutamate-induced IP3 production in response to neural activity. These fluxes are expressed as follows:

$$J_{ER} = c_1 v_1 [Ca^{2+}]^3 h^3 [IP_3]^3 \frac{\left(c_0/c_1 - \left(1 + \frac{1}{c_1}\right)[Ca^{2+}]\right)}{(([IP_3] + d_1)([Ca^{2+}] + d_5))^3};$$

$$J_{pump} = \frac{v_3 [Ca^{2+}]^2}{k_3^2 + [Ca^{2+}]^2};$$

$$J_{leak} = c_1 v_2 (c_0/c_1 - (1 + 1/c_1)[Ca^{2+}]); \qquad (9)$$

$$J_{in} = \frac{v_6 [IP_3]^2}{k_2^2 + [IP_3]^2};$$

$$J_{out} = k_1 [Ca^{2+}];$$

$$J_{PLC\delta} = \frac{v_4([Ca^{2+}] + (1-\alpha)k_4)}{[Ca^{2+}] + k_4}.$$

Astrocytes interact with each other through gap junctions. Gap junctions are permeable to the second messenger IP3 and to calcium ions[46,47]. Currents $J_{Gcam}$ and $J_{Gip3m}$ describe the diffusion of $Ca^{2+}$ ions and IP3 molecules via gap junctions of the *m*th astrocyte and can be expressed as follows:

$$J_{Gcam} = d_{ca} \sum_j ([Ca^{2+}]_j - [Ca^{2+}]_m);$$

$$J_{Gip3m} = d_{ip3} \sum_j ([IP_3]_j - [IP_3]_m), \tag{10}$$

where *j*, $d_{ca}$ and $d_{ip3}$ represent, respectively, the number of astrocytes connected to the *m*th astrocyte and the $Ca^{2+}$ and IP3 diffusion rates. Biophysical meaning of all parameters in **Eqs. 8-10** and their values can be found in the article[31] and in **Table 2** (astrocytic network parameters).

**Bidirectional neuron-astrocyte interaction.** The amount of neurotransmitter-glutamate that diffuses from the synaptic cleft associated with the *i*th pyramidal neuron and reaches the astrocyte is described by the following equation[20,48]:

$$\frac{dG_i}{dt} = -\alpha_{glu}G_i + k_{glu}H(V_i - 30mV), \tag{11}$$

where $\alpha_{glu}$ is the glutamate clearance constant, $k_{glu}$ is the release efficiency, *H* is the Heaviside step function, and $V_i$ is the membrane potential of *i*th pyramidal neuron. Glutamate contacts the mGluRs on the astrocyte membrane and initiates the production of IP3. The flux $J_{glu}$ represents the glutamate-induced IP3 production and is defined as follows:

$$J_{glu} = \begin{cases} A_{glu}, & \text{if } t_0 < t \leq t_0 + t_{glu}, \\ 0, & \text{otherwise}; \end{cases} \tag{12}$$

here $t_0$ represents the moment when the total level of glutamate concentration in all synapses associated with this astrocyte reaches a threshold:

$$\left(\frac{1}{N_{AE}} \sum_{i \in N_{AE}} [G_i \geq G_{thr}]\right) \geq F_{act}, \tag{13}$$

where the parameter $G_{thr}=0.2$ is the threshold for glutamate, [x] is the Iverson bracket. $F_{act} = 0.75$ denotes the fraction of synchronously spiking neurons of the neuronal ensemble corresponding to the astrocyte.

Experimental studies have shown that astrocytes are able to facilitate synaptic transmission due to the action of glutamate released from astrocytes. More precisely, we consider that the astrocytic glutamate induces potentiation of the excitatory synapse via NMDAR-dependent postsynaptic slow inward currents (SICs) generation[49,50] and mGluR-dependent heterosynaptic facilitation of presynaptic glutamate release[51,52,53]. In the SNAN, we propose that $Ca^{2+}$ elevation in astrocytes results in glutamate release, which can modulate the synaptic strength of all synapses corresponding to the morphological territory of a given astrocyte. For simplicity, astrocyte-induced enhancement of synaptic weight of the affected excitatory synapses, $\overline{w_{synEE}}$, is described as follows:

$$\overline{w_{synEE}} = w_{synEE}(1 + v_{Ca}), w_{synEE} \in [0, w_{synEEmax}];$$
$$v_{Ca} = v_{Ca}^* H([Ca^{2+}]_m - [Ca^{2+}]_{thr}), \quad (14)$$

where $w_{synEE}$ is the weight of the excitatory synapse trained according to Hebb's rule, $v_{Ca}^* = 2$ represents the strength of the astrocytic modulation of the synaptic weight, $H(x)$ is the Heaviside function, $[Ca^{2+}]_{thr}$ denotes the threshold $Ca^{2+}$ concentration in the astrocyte *m*. The feedback from the astrocytes to the neurons is activated when the astrocytic $Ca^{2+}$ concentration is larger than $[Ca^{2+}]_{thr}$, and the fraction of synchronously spiking neurons of neuronal ensemble corresponding to the astrocyte $F_{astro}$ during the period of $\tau_{syn} = 5$ ms. The duration of astrocyte-induced enhancement of synaptic transmission is fixed and equal to $\tau_{astro} = 20$ ms.

Model equations are integrated using the Runge-Kutta fourth-order method with a fixed time step, $\Delta t = 0.1$ ms. A detailed listing of model parameters and values can be found in **Methods** in **Tables 1** (neural network model), **2** (astrocytic network parameters), **3** (neuron-astrocytic interaction parameters) and **4** (training and testing protocol parameters).

**Memory performance metrics.** To measure memory performance of the proposed SNAN, we calculate the correlation of recalled pattern with the ideal sample in the following way:

$$M_i(t) = \left[ \left( \sum_{k=t-w}^{t} [V_i(k) > V_{thr}] \right) > 0 \right],$$

$$C(t) = \frac{1}{2} \left( \frac{1}{|P|} \sum_{i \in P} M_i(t) + \frac{1}{W \cdot H - |P|} \sum_{i \notin P} (1 - M_i(t)) \right), \quad (15)$$

$$C_P = \frac{1}{|T_P|} \max_{t \in T_P} C(t);$$

here $w = 10$ frames $= 1$ ms, $P$ - a set of pixels belonging ideal pattern, $W$, $H$ - network dimensions, $V_{thr}$ - spike threshold, $I$ - indicator function, $T_P$ - a set of frames in the tracking range of pattern $P$. In a sense, this correlation metric can be associated with 1 - $d$ averaged between pattern and background, where $d$ is the Hamming distance.

## Table 1 Neural network parameters[42,54]

| Parameter | Parameter description | Value |
|---|---|---|
| $W \times H$ | pyramidal neurons layer grid size | 79×79 |
| $W_1 \times H_1$ | interneurons layer grid size | 40×40 |
| $a$ | time scale of the recovery variable | 0.1 |
| $b$ | sensitivity of the recovery variable to the sub-threshold fluctuations of the membrane potential | 0.2 |
| $c$ | after-spike reset value of the membrane potential | -65mV |
| $d$ | after-spike reset value of the recovery variable | 2 |
| $\eta$ | synaptic weight without astrocytic influence | 0.025 |
| $E_{syn}$ | synaptic reversal potential for excitatory synapses<br>synaptic reversal potential for inhibitory synapses | 0 mV<br>-90 mV |
| $k_{syn}$ | slope of the synaptic activation function | 0.2 mV |
| **Connections within a pyramidal neurons layer:** | | |
| $N_{outEE}$ | number of output connections per each neuron | 200 |
| $\lambda_{EE}$ | rate of the exponential distribution of synaptic connections distance | 15 |
| $g_{synEE}$ | change in the value of the weight during training | 0.007 |
| $w_{synEEmax}$ | maximum synaptic weight | 0.05 |
| **Connections from a pyramidal neurons layer to interneurons layer:** | | |
| $N_{outEI}$ | number of output connections per each neuro | 5 |
| $\lambda_{EI}$ | rate of the exponential distribution of synaptic connections distance | 2 |

| Parameter | Parameter description | Value |
|---|---|---|
| $w_{synEI}$ | weight of synaptic connections | 0.1 |
| **Connections from a interneurons layer to pyramidal neurons layer:** | | |
| $N_{outIE}$ | number of output connections per each neuron | 2000 |
| $\lambda_{IE}$ | rate of the exponential distribution of synaptic connections distance | 80 |
| $g_{synEE}$ | change in the value of the weight during training | 0.007 |
| $w_{synIEmax}$ | maximum synaptic weight | 0.05 |
| $f^*$ | the threshold firing rate of the pyramidal neuron for training of IE connections | 0.3 |

### Table 2 Astrocytic network parameters[31]

| Parameter | Parameter description | Value |
|---|---|---|
| $M \times N$ | astrocytic network grid size | 26×26 |
| $c_0$ | total $Ca^{2+}$ in terms of cytosolic vol | 2.0 μM |
| $c_1$ | (ER vol)/(cytosolic vol) | 0.185 |
| $v_1$ | max $Ca^{2+}$ channel flux | 6 s$^{-1}$ |
| $v_2$ | $Ca^{2+}$ leak flux constant | 0.11 s$^{-1}$ |
| $v_3$ | max $Ca^{2+}$ uptake | 2.2 μM s$^{-1}$ |
| $v_6$ | maximum rate of activation dependent calcium influx | 0.2 μM s$^{-1}$ |
| $k_1$ | rate constant of calcium extrusion | 0.5 s$^{-1}$ |
| $k_2$ | half-saturation constant for agonist-dependent calcium entry | 1 μM |
| $k_3$ | activation constant for ATP- $Ca^{2+}$ pump | 0.1 μM |
| $d_1$ | dissociation constant for $IP_3$ | 0.13 μM |
| $d_2$ | dissociation constant for $Ca^{2+}$ inhibition | 1.049 μM |
| $d_3$ | receptor dissociation constant for $IP_3$ | 943.4 nM |
| $d_5$ | $Ca^{2+}$ activation constant | 82 nM |
| $\alpha$ | | 0.8 |
| $v_4$ | max rate of $IP_3$ production | 0.3 μM s$^{-1}$ |
| $1/\tau_r$ | rate constant for loss of $IP_3$ | 0.14 s$^{-1}$ |
| $IP_3^*$ | steady state concentration of $IP_3$ | 0.16 μM |
| $k_4$ | dissociation constant for $Ca^{2+}$ stimulation of $IP_3$ production | 1.1 μM |
| $d_{ca}$ | $Ca^{2+}$ diffusion rate | 0.05 s$^{-1}$ |
| $d_{ip3}$ | $IP_3$ diffusion rate | 0.05 s$^{-1}$ |

## Table 3 Neuron-astrocytic interaction parameters[48]

| Parameter | Parameter description | Value |
|---|---|---|
| $N_{AE}$ | number of neurons interacting with one astrocyte | 16, 4×4 |
| $α_{glu}$ | glutamate clearance constant | 50 s$^{-1}$ |
| $k_{glu}$ | efficacy of the glutamate release | 600 μM s$^{-1}$ |
| $A_{glu}$ | rate of IP$_3$ production through glutamate | 5 μM s$^{-1}$ |
| $t_{glu}$ | duration of IP$_3$ production through glutamate | 60 ms |
| $G_{thr}$ | threshold concentration of glutamate for IP$_3$ production | 0.2 |
| $F_{act}$ | fraction of synchronously spiking neurons required for the emergence of Ca$^{2+}$ elevation | 0.75 |
| $F_{astro}$ | fraction of synchronously spiking neurons required for the emergence of astrocytic modulation of synaptic transmission | 0.5 |
| $v_{Ca}^{*}$ | strength of the astrocyte-induced modulation of synaptic weight | 2 |
| $[Ca^{2+}]_{thr}$ | threshold concentration of Ca$^{2+}$ for the astrocytic modulation of synapse | 0.15 μM |
| $τ_{astro}$ | duration of the astrocyte-induced modulation of synapse | 20 ms |

## Table 4 Training and testing protocol parameters

| Parameter | Parameter description | Value |
|---|---|---|
| $A_{stim}$ | stimulation amplitude in training | 80 μA |
| $t_{stim}$ | stimulation duration in training | 2 ms |
| $t_{between\ stim}$ | time between patterns in training | 3 ms |
|  | noise level in training | 5% |
| $A_{test}$ | stimulation amplitude in test | 8 μA |
| $t_{test}$ | stimulation length in test | 20 ms |
| $t_{between\ test}$ | time between patterns in test | 50 ms |
|  | noise level in test | 20% |
| $P$ | number of pre-training patterns | 40; 20 |

## Code availability

The code is available at https://github.com/altergot/Neuron-astrocyte-network-Situation-associated-memory.

## Acknowledgements

The reported study was funded by the Ministry of Science and Higher Education of the Russian Federation (project no. 075-15-2020-808).

## Author contributions

S.Y.G. – Design of model experiments, model simulation, data analysis, results interpretation, writing the manuscript.

Y.A.T. – Model simulation, data analysis, results interpretation, writing the manuscript.

M.I.K. – Model simulation, data analysis, results interpretation, writing the manuscript.

I.Y.T. – Results interpretation, writing the manuscript.

V. B. K. – Data analysis, results interpretation, writing the manuscript.

A. A. Z. – Design of model experiments, model formulation, results interpretation, writing the manuscript.

A.N.G. – Design of model experiments, model formulation, results interpretation, writing the manuscript.

## Competing interests

The authors declare no competing interests.